\documentstyle[aps,preprint]{revtex}
\mathchardef\step="1376
\begin{document}
\draft

\title{
\centerline{\bf GENERATING NEW MAGNETIC UNIVERSE}
\centerline{\bf SOLUTIONS FROM OLD}
}
\author{David Garfinkle}
\address{
\centerline{Department of Physics, Oakland University,
Rochester, MI 48309}}

\author{M. A. Melvin}
\address{
\centerline{1300 Orchid Drive, Santa Barbara, CA 93111}
}
\maketitle

\null\vspace{-1.75mm}

\begin{abstract}
In this paper we apply the techniques which have been developed over the last
few decades for generating nontrivially new solutions of the Einstein-Maxwell
equations from seed solutions for simple spacetimes.  The simple seed spacetime
which we choose is the ``magnetic universe'' to which we apply the Ehlers
transformation.  Three interesting non-singular metrics are generated.  Two
of these may be described as ``rotating magnetic universes'' and the third as
an ``evolving magnetic universe.''  Each is causally complete $-$ in that
all timelike and
lightlike geodesics do not end in a finite time or affine parameter.  We also
give the electromagnetic field in each case.  For the two rotating stationary
cases we give the projection with respect to a stationary observer of the
electromagnetic field into electric and magnetic components.
\end{abstract}

\narrowtext

\section{INTRODUCTION}

Begining in the late 50's and continuing into the 80's an interesting technique
was developed for generating new vacuum, and Einstein-Maxwell, solutions from
a given ``seed'' solution having at least one Killing vector.  This was of
special interest when the Killing vector field corresponds to an azimuthal
symmetry about an axis.  Even more fruitful was the case when the seed metric
had stationary axial symmetry about an axis with two corresponding Killing
vector fields,
$ {\bf \partial / \partial \phi }$
and
$ {\bf \partial / \partial t }$ .
The technique, as we shall apply it, starts with the differential form
$ K $
corresponding to a given Killing vector field
$ {\bf K} $.
This Killing form in combination with forms associated with the self-dual
part of the Maxwell field $-$ when present in the seed
spacetime $-$ allows one to define succesively
two scalar potentials $-$ one the Killing-Maxwell potential
$ \Phi $
$-$
and the other the Killing-Einstein potential
$ \cal E $
defined by Ernst.
Certain transformations of the scalar potentials of the seed metric
lead to nontrivially new potentials, a new Killing form and a new tetrad basis
defining a new metric.

The transformations from old to new solutions were found piecemeal by various
investigators.  One of the earliest, given by J. Ehlers, \cite{ehlers}
is the one we shall apply in the present paper.  Others were found by
B. K. Harrison\cite{harrison} and R. Geroch.\cite{geroch}  Finally all
together were codified in a group structure by W. Kinnersley.\cite{kinnersley}
They have also been investigated extensively first by F. Ernst and then
by I. Hauser and F. Ernst,\cite{hauern} whose procedure we follow here.

It is to be noted that even when a seed solution is well behaved-even
altogether free from singularities-the resulting solutions often are not well
behaved,
and therefore not of great physical interest.  In this paper we apply the
Ehlers transformation using the magnetic universe\cite{melvin} as a seed
solution.  The magnetic universe is globally well behaved with no
singularities.  In the magnetic universe there are three Killing vector
fields which, applied individually, allow the generation of three distinct
new spacetimes.  Each of these spacetimes is timelike and lightlike
geodesically complete (nonsingular).  Two of them are
stationary-axisymmetric representing
rotating magnetic universes.  The third is cylindrically symmetric
but nonstationary $-$ evolving in time.

In section II we describe the seed metric - the magnetic universe.  We
then describe the derivation of the potentials, and the
Ehlers transformation.  In section III we apply the Ehlers transformation to
obtain the three new metrics.
In section IV we examine global structure and demonstrate that each spacetime
is nonsingular.  We also discuss the properties of the electromagnetic fields
of each solution.  Section V contains our conclusions.

\section{SEED METRIC AND TRANSFORMATION}

Among configurations of matter or field which are in static equilibrium
under their own gravitational attraction one of the simplest is a parallel
bundle of magnetic flux.  Associated with the equilibrium magnetic field
distribution is a well determined geometry with cylindrical
symmetry.\cite{melvin}
The field distribution together with its associated geometry has come
to be called ``the magnetic universe.''  The geometry of the
cylindrical magnetic universe is given by
$$
d {s^2} = {v^2} \, \left (  - \, d {t^2} \, + \, d {z^2} \,
+ \, d {\rho ^2}   \right ) \; + \;
{{\rho ^2} \over {v^2}}  \, d {\phi ^2} = {e^z} \bullet {e^z} \;
+ \;  {e^\phi } \bullet {e^\phi } \; + \;  {e^\rho } \bullet
{e^\rho } \; - \;  {e^t} \bullet {e^t} \;  \; \; .
\eqno({\rm II}.1)
$$
where
$ z , \, \phi , \, \rho $
and
$ t $
are the usual cylindrical spacetime coordinates.
Here
$ v $
is given by
$ v \equiv 1 \, + \, {\rho ^2} $,
and an orthonormal basis of one-forms is
$$
{e^z} \equiv v d z \; \; , \; \; {e^\phi} \equiv {\rho \over v} \,
d \phi \; \; , \; \; {e^\rho} \equiv v d \rho \; \; , \; \;
{e^t} \equiv v d t \; \; .
\eqno({\rm II}.2)
$$
The volume form is
$ \epsilon = {v^2} \rho dt \wedge dz \wedge d \rho \wedge d \phi $.
We use a notation where vectors are written in boldface type and forms in
italic type.
The Maxwell field, which varies only with radius
$ \rho $,
is given by
$$
F = {{2 \rho} \over {v^2}} \, d \rho \wedge d \phi = - \,
d \left ( {1 \over v} \right ) \wedge d \phi =
{2 \over {v^2}} \; {e^\rho } \wedge {e^\phi} \equiv
{2 \over {v^2}} \; {e^{\rho \phi } }
\eqno({\rm II}.3)
$$
where we define the symbol
$ {e^{ab}} \equiv {e^a} \wedge {e^b} $.
The coordinates and metric are dimensionless because we are measuring all
lengths in units of the ``range radius''
$ {\bar a} = 2/ {B_0} $
(gravitational length units) and where
$ B_0 $
is the magnitude of the magnetic field on the axis, (measured in gauss).  In
cgs units
$$
{\bar a} = {{2 {c^2}} \over {{B_0} {G^{1/2}}}} = {{1.96} \over {B_0}} \;
\times {{10}^{24}} {\rm cm} \; \; \; .
$$
The three Killing vector fields are
$ {\bf \partial /\partial z \; \; , \; \; \partial /
\partial \phi \; \; {\rm and}
\partial /\partial t }$
with corresponding one-forms
$  {v^2} d z \; \; , \; \; {{(\rho /v)}^2} d \phi \; \; {\rm and} - {v^2} d t$.
The three Killing fields give rise to three conserved quantities for the
geodesic equation.  These quantities have been used to solve the
geodesic equation in closed form.\cite{melvin2}

We now discuss the solution generating technique.  Starting with a solution of
the Einstein-Maxwell equations with a Killing field,
one produces two scalar potentials
$ \Phi $
and
$ \cal E $.
One then uses these potentials with a solution generating
technique ({\it e.g.} the Ehlers transformation) to find an
orthonormal tetrad of one-forms for
the new metric.

First we introduce some notation, the so called ``step product''
$ M \step \; N $
of two forms
which leads to a contracted form.  This is only different from zero when
the degree of the form
$ M $
on the left is not greater than the degree of
$ N $.
The step product of two 1-forms is the dot product.  Given a one-form
$ K $
and a two-form
$ Q $
which is the wedge product of two 1-forms,
$ Q \equiv u \wedge v $,
we define the step product:
$$
K \step \; Q \equiv K \step \; \left ( u \wedge v \right ) = u
\left ( K \cdot v \right ) \; - \; v \left ( K \cdot u \right )
\eqno({\rm II}.4)
$$
where the dot represents the usual inner product of two fields $-$
the same for vectors and the corresponding 1-forms.

Now let
$W$
be a self-dual 2-form (where the dual of a two form
$F$
is
$- i {^*F}$).  Then one can show that
$$
2 \times {\rm self \; dual \;  of \; } \left [ K \wedge (K \step \; W)
\right ] =  - \; (K \cdot K) W
\eqno({\rm II}.5)
$$
This relation will be particularly useful in enabling us to reconstruct
the electromagnetic field from a Killing-dependent scalar Maxwell potential
$ \Phi $ which we shall now define.

{}From the Maxwell two-form we define the
self-dual Maxwell two-form
$ W^M $
by
$$
{W^M} \equiv F \, - \, i \, {^*F} \; \; \; .
\eqno({\rm II}.6)
$$
For example, in the special case of the magnetic universe
$$
{W^M} = {2 \over {v^2}} \; \left ( {e^{\rho \phi}} \, - \, i {e^{t z}} \right )
= {{2 \rho } \over {v^2}} \; d \rho \wedge d \phi \; - \;
2 i d t \wedge d z \; \; \; .
\eqno({\rm II}.7)
$$

Now we define what we may call the ``Killing-Maxwell 1-form''
$ K \step \; {W^M} $
and the complex scalar Maxwell potential
$ \Phi $
by
$$
d \Phi = K \step \; {W^M} \; \; \; .
\eqno({\rm II}.8)
$$
The integrability conditions for this equation are satisfied
as a consequence of the source-free Maxwell equations,
provided the Lie derivative of
$ F $
vanishes along the associated non-null Killing field; {\it i.e.},
the electromagnetic field is constant along the Killing trajectories.
The integrability is shown as follows: we have, for any 2-form
$ W $,
the expression of the Lie derivative in terms of the exterior
derivative and the step product:\cite{ref8}
$$
{{\cal L}_{\bf K}} W =  K \step \;  d W  \; - \; d \left (
K \step \; W \right ) \; \; \; .
$$
Here, in our case,
$ W^M $
is linear and homogeneous in
$ F $
and
$ ^*F $
and $-$ with
$ d {W^M} = d F = d {^*F} = 0 $
as well as
$  {{\cal L}_{\bf K}} F = {{\cal L}_{\bf K}} {^*F} = {{\cal L}_{\bf K}} {W^M}
= 0 - $
it follows that
$ d \left ( K \step \; {W^M} \right ) = 0 $.

Next we define the self dual Einstein-Maxwell-Harrison-Ernst two-form
$ W^E $
by
$$
{W^E} \equiv - \; \left ( d K \, + \, 2 {\bar \Phi} F \right ) \; + \;
 i \left ( {^*d K} \, + \, 2 {\bar \Phi} {^*F} \right )
\eqno({\rm II}.9)
$$
where
$ \bar \Phi $
is the complex conjugate of
$ \Phi $.
Then we define the Killing-Einstein 1-form
$ K \step \; {W^E} $
and the scalar Ernst potential
$ \cal E $
by
$$
d {\cal E} = K \step \; {W^E} \; \; \; .
\eqno({\rm II}.10)
$$
The integrability conditions for this equation are satisfied as
a consequence of the Einstein-Maxwell equations using the
vanishing Killing Lie derivative conditions analogous to
before.  The constant in
$ \cal E $
is fixed by requiring
$$
{\rm Re} {\cal E} = f - {{|\Phi |}^2}
\eqno({\rm II}.11)
$$
where
$ f $
is minus the norm of the Killing field, {\it i.e.}
$ f \equiv - {\bf K} \cdot {\bf K} $.

For a hypersurface orthogonal Killing vector
$ {\bf K }$
a useful relation is found by substituting equation (II.9) into
equation (II.10) giving
$$
d {\cal E} = d f \; - \; d {\bar \Phi } d {\Phi} \; \; \; .
$$
This implies
$$
d {\rm Im} {\cal E} = i \left ( {\bar \Phi } d \Phi \; - \; \Phi d
{\bar \Phi } \right ) \; \; \; .
\eqno({\rm II}.12)
$$
Thus $-$ as is true in each of our 3 cases $-$ if
$\Phi$
is pure real or pure imaginary,
${\rm Im}  {\cal E} $
is a constant which may be taken as zero.  In more general situations
${\rm Im}  {\cal E} $
will be determined by equation (II.12).

We will also need only the one-forms
$ A^E $
and
$ M^{EE} $
defined by
$$
d {A^E} = {W^E}
\eqno({\rm II}.13)
$$
and
$$
d {M^{EE}} = 2 {\rm Re} \left ( {\bar {\cal E}} {W^E} \right ) \; \; \; .
\eqno({\rm II}.14)
$$
These form-potentials are chosen to satisfy first the gauge condition
$ K \step \; {A^E} = {\cal E} $
and, consistent with this, we have the further gauge condition
$ K \step \; {M^{EE}} = {{| {\cal E} |}^2} $.
The consistency can be seen from applying the
$ K $
contraction to equation (II.14) and noting that $-$ with the Lie derivatives of
$ M^{EE} $
and
$ A^E $
vanishing $-$ the
$ K $
contraction commutes with the exterior derivative and we have
$$
d \left ( K \step \; {M^{EE}} \right ) = 2 {\rm Re} \left [
{\bar {\cal E}} d \left ( K \step \; {A^E} \right ) \right ] =
2 {\rm Re} \left ( {\bar {\cal E}} d {\cal E} \right ) = d \left (
{{\left | {\cal E} \right | }^2} \right ) \; \; \; .
$$

We are now ready to discuss the Ehlers transformation.  Let
$ \beta $
be a real number.  Define the scalar
$ \Lambda $
by
$$
\Lambda \equiv 1 \, + \, i \beta {\cal E} \; \; \; .
$$
The Ehlers transformation, expressed in the group formalism developed
by Kinnersley\cite{kinnersley} is:
$$
\left ( \matrix {1\cr {{\cal E} '} \cr {\Phi '}\cr}\right ) =
{1 \over \Lambda} \; \left ( \matrix {1&i \beta &0\cr 0&1&0\cr
0&0&1\cr}\right ) \left ( \matrix {1\cr {\cal E} \cr \Phi \cr}\right )
= \left ( \matrix {1\cr {\cal E}/\Lambda  \cr \Phi /\Lambda \cr}\right )
$$
which implies for the norm of the transformed Killing field in the new
spacetime
$$
{f '} = {\rm Re} {{\cal E} '} \; + \; {{\left | {\Phi '} \right |}^2}
= {{{\rm Re} {\cal E} \, + \, {{|\Phi |}^2}} \over {{|\Lambda |}^2}}
= {f \over {{|\Lambda |}^2}} \; \; \; .
\eqno({\rm II}.15)
$$
The more general formulas given in reference\cite{hauern} simplify in the case
of the Ehlers transformation.  It can be shown that the transform
$ K ' $
of the Killing form
$ K $
satisfies the equation
$$
{{f '}^{- 1}} {K '} = {f^{- 1}} K {{|\Lambda |}^2} \; + \; {\beta ^2}
{M^{EE}} \; - \; 2 \beta {\rm Im} {A^E} \; \; \; .
\eqno({\rm II}.16)
$$
In all three cases that occur with the Killing vectors of the
magnetic universe, which we will develop in the following,
$ \cal E $
is purely real and
$ {M^{EE}} = {{\cal E}^2} d {x^a} $,
where
$ x^a $
is the coordinate for the relevant Killing field.  The vector field is
$ \partial / \partial {x^a} $
and the form field is
$$
{K^a} = - f d {x^a} = - \; {f \over {\sqrt {|f|}}} \;  {e^a}
\eqno({\rm II}.17)
$$
($ f < 0 $
for
$ x^a $
spacelike;
$ f > 0 $
for
$ x^a $
timelike).  Equation (II.16) then simplifies to
$$
{{f '}^{- 1}} {{K^a} '} = - \; d {x^a} \;  {{|\Lambda |}^2} \; + \;
{\beta ^2} {{\cal E}^2} d {x^a} \; - \; 2 \beta {\rm Im} {A^E}
$$
which, upon using
$ {{|\Lambda |}^2} = 1 + {\beta ^2} {{\cal E}^2} $
can be written
$$
{{f '}^{- 1}} {{K^a} '} = - \, \left ( d {x^a} \; + \; 2 \beta {\rm Im}
{A^E} \right ) \, = - \, \left ( {1 \over {\sqrt {|f|}}} \;  {e^a}  \;
+ \; 2 \beta {\rm Im} {A^E} \right ) \; \; \; .
\eqno({\rm II}.18)
$$
Now a convenient tetrad in the new spacetime can be constructed from
the tetrad used to describe the original spacetime.
The three tetrad forms orthogonal to
$ K $
transform as follows:
$$
{e^a} \to {e^{a '}} = | \Lambda | {e^a}  \; \; \; . \; \; \;
\; \; \left ( {e^a} \perp K \right )
\eqno({\rm II}.19)
$$
The remaining tetrad form which we shall designate
$ e^k $
(${e^k} \parallel K $) transforms as
$$
{e^a} \to  {e^{a '}} = {{\sqrt {|f|}} \over {| \Lambda |}} \;
\left ( d {x^a} \; + \; 2 \beta {\rm Im} {A^E} \right ) = - \;
{{\sqrt {|f|}} \over {|\Lambda |}} \; {{f ' }^{- 1}} {K '} \; \; \; \; .
\eqno({\rm II}.20)
$$

The Maxwell scalar potential transforms as
$$
{\Phi '} = {\Phi \over {1 \, + \, i \beta {\cal E}}} =
{{\Phi \left ( 1 \, - \, i \beta {\cal E} \right ) } \over {{|\Lambda |}^2}}
\eqno({\rm II}.21)
$$
and equation (II.5) yields
$$
W \to {W '} = {{f ' }^{- 1}} \left ( {K '} \wedge \left [
{K '} \step \;  {W '} \right ] \; - \; i {^*} \left [ {K '} \wedge
\left ( {K '} \step \; {W '} \right ) \right ] \right )
$$
$$
= {{f ' }^{- 1}} \left ( {K '} \wedge d {\Phi '} \; - \; i {^*} \left [
{K '} \wedge d {\Phi '} \right ] \right )
\eqno({\rm II}.22)
$$
or, separating out the real part on both sides of the equation, we have
$$
F \to {F '} =  {{f '}^{- 1}} {K '} \wedge d \left ( {\rm Re}
{\Phi '} \right ) \; + \; * \left [ {{f '}^{- 1}} {K '} \wedge
d \left ( {\rm Im} {\Phi '} \right )   \right ]
\eqno({\rm II}.23)
$$
$$
= - \; {{|\Lambda |} \over {\sqrt {|f|}}} \; \left (  {e^{k '}} \wedge
d \left ( {\rm Re} {\Phi '} \right ) \; + \; {^* } \left [  {e^{k '}}
\wedge d \left ( {\rm Im} {\Phi '} \right )   \right ] \right )\; \; \; .
$$
In the two types of cases of interest to us here: 1. the seed
$ \Phi $
pure real and equal to
$ {\cal E} $
(the case of the
$ \phi $
metric); 2. the seed
$ \Phi $
pure imaginary,
$ {\cal E} $
real (the case of the
$ t $
and
$ z $
metrics) the general equation reduces as follows:
$$
\Phi \; {\rm real} = {\cal E} : \; \; \; {F '} = - \; {{|\Lambda |}
\over {{{|\Lambda |}^4} {\sqrt {| f|}}}} \; \left [ \left (
1 \, - \, {\beta ^2} {{\cal E}^2} \right )
{e^{k '}} \wedge d \Phi \; - \; 2 \beta \Phi {^*} \left (
{e^{k '}} \wedge d \Phi \right ) \right ] \; \; \; ,
$$
$$
\Phi \; {\rm imaginary}, \; \; {\cal E} \; {\rm real} : \; \; \;
{F '} = - \; {{|\Lambda |} \over {{{|\Lambda |}^4}
{\sqrt {| f|}}}} \; \left ( \beta \left [ {\cal E} \left (
1 \, + \, {\beta ^2} {{\cal E}^2} \right ) {e^{k '}} \wedge d
|\Phi | \; + \; |\Phi | \left ( 1 \, - \, {\beta ^2} {{\cal E}^2} \right )
{e^{k '}} \wedge d {\cal E} \right ] \right .
$$
$$
\left . + \; {^*} \left [ \left ( 1 \, + \, {\beta ^2} {{\cal E}^2}
\right ) {e^{k '}} \wedge d |\Phi |\; - \; 2 {\beta ^2}
|\Phi | {\cal E} {e^{k '}} \wedge d {\cal E} \right ] \right )
$$
Choosing an observer with four-velocity
$ {\bf u} $
the Maxwell tensor can be decomposed into the electric and magnetic fields
measured by this observer.  The electric and magnetic fields are given by
$$
E = u \step \;  F \; \; \; ,
$$
$$
B = - \, u \step \; {^* F} \; \; \; .
$$
where
$ u $
is the 1-form corresponding to the vector
$ {\bf u} $.
In our case we choose
$ u $
to be minus the timelike tetrad form.

\section{THREE NEW SPACETIMES}

We now apply the Ehlers transformation to the magnetic universe to produce new
spacetimes.  We get a different spacetime for each of the Killing vectors of
the magnetic universe.

\subsection{The $\phi$ metric}

The two-form
$ W^M $
for the magnetic universe is given by
$$
{W^M} = {{2 \rho} \over {v^2}} \; d \rho \wedge d \phi \;
- \; 2 i d t \wedge d z \; \; \; .
\eqno({\rm III}.1)
$$
Using the Killing vector
$ \bf K = \partial / \partial \phi $
it follows that
$$
K \step \; {W^M} =  {{2 \rho } \over {v^2}} \; d \rho \; \; \; .
$$
and therefore that
$ \Phi = - \, {v^{- 1}} $.
We then find that
$ W^E $
is given by
$$
{W^E} = {{2 \rho} \over {v^2}} d \rho \wedge d \phi \;
- \; 2 i d z \wedge d t \; \; \; .
\eqno({\rm III}.2)
$$
Using equation (II.10) we then find
$$
{\cal E} = - {v^{-1}} \; \; \; .
$$
We then find, using equations (II.9) and (II.13) that
$$
{A^E} = - {v^{-1}} d \phi \; - \; 2 i z d t \; \; \; .
$$
Then using equations (II.10) and (II.14) we find
$$
{M^{EE}} = {v^{- 2}} \, d \phi \; \; \; .
$$
The new tetrad is given as follows:
$$
{e^{1 '}} = {{\left ( {v^2} \, + \, {\beta ^2} \right )}^{1/2}} \; d z
\; \; \; ,
$$
$$
{e^{2 '}} = \rho {{\left ( {v^2} \, + \, {\beta ^2} \right )}^{-1/2}} \;
\left [ d \phi \, + \, 4 \beta z d t \right ] \; \; \; ,
$$
$$
{e^{3 '}} = {{\left ( {v^2} \, + \, {\beta ^2} \right )}^{1/2}} \;
d \rho \; \; \; ,
$$
$$
{e^{4 '}} = {{\left ( {v^2} \, + \, {\beta ^2} \right )}^{1/2}} \;
d t \; \; \; .
$$
Thus the metric is given by
$$
d {s^2} = \left ( {v^2} \, + \, {\beta ^2} \right ) \; \left [ - d {t^2} \; +
\; d {z^2} \; + \; d {\rho ^2}  \right ] \; + \; {\rho ^2}
{{\left ( {v^2} \, + \, {\beta ^2}  \right )}^{-1}}
\; {{\left [ d \phi \; + \; 4 \beta z d t \right ] }^2} \; \; \; .
\eqno({\rm III}.3)
$$

The new Maxwell tensor is given in terms of the new tetrad by
$$
{F '} = 2 \; {{{v^2} - {\beta ^2}} \over {{\left ( {v^2} +
{\beta ^2} \right ) }^2}} \; {e^{3 '}} \wedge {e^{2 '}} \;
+ \; {{4 \beta v} \over {{\left ( {v^2} + {\beta ^2} \right ) }^2}} \;
{e^{4 '}} \wedge {e^{1 '}} \; \; \; .
\eqno({\rm III}.4)
$$
The corresponding electric and magnetic fields are
$$
E = - \;  {{4 \beta v} \over {{\left ( {v^2} + {\beta ^2} \right ) }^2}} \;
{e^{1 '}} \; \; \; ,
$$
$$
B = 2 \;  {{{v^2} - {\beta ^2}} \over {{\left ( {v^2}
+ {\beta ^2} \right ) }^2}} \; {e^{1 '}} \; \; \; .
\eqno({\rm III}.5)
$$
The electric and magnetic fields point in the (plus or minus)
$ z $
direction.  Note that in the
$ \beta \to 0 $
limit the metric and Maxwell tensor reduce to those of the
magnetic universe.  Thus the parameter
$ \beta $
can be regarded as the magnitude of the rotation.

\subsection{The $t$ metric}

Using the Killing vector
$ \bf K = \partial / \partial t $
it follows that
$$
K \step \; {W^M} = 2 i d z \; \; \; , \; \; \; \;  \Phi = 2 i z  \; \; \; .
\eqno({\rm III}.6)
$$
It then follows from equation (II.9) that
$$
{W^E} = d \left ( {v^2} \, - \, 4 {z^2} \right ) \wedge d t \;
+ \; 4 i d \left ( {\rho ^2} {v^{-1}} z \right ) \wedge d \phi \; \; \; .
$$
It then follows from equation (II.10) that
$$
{\cal E} = {v^2} \, - \, 4 {z^2} \; \; \; .
\eqno({\rm III}.7)
$$
Now using equations (II.9) and (II.13) we find
$$
{A^E} = {\cal E} d t \; + \; 4 i {\rho ^2} {v^{-1}} z d \phi \; \; \; .
\eqno({\rm III}.8)
$$
Similarly using equations (III.7) and (II.14) we find again that
$$
{M^{EE}} = {{\cal E}^2} d t \; \; \; .
$$
We are now ready to compute the new tetrad.  Since
$ \Lambda = 1 \, + \, i \beta {\cal E} $
and here
$ |\Lambda| = {{\left [ 1 \, + \, {\beta ^2} {{\left ( {v^2} \,
- \, 4 {z^2}\right ) }^2} \right ] }^{1/2}} $
we find that three of the tetrad vectors are given by
$$
{e^{1 '}} = |\Lambda | \; v d z \; \; \; ,
$$
$$
{e^{2 '}} =   |\Lambda | \; \rho {v^{-1}} d \phi \; \; \; .
$$
$$
{e^{3 '}} = |\Lambda | \; v d \rho \; \; \; ,
$$
Finally using equations (II.20) and (III.8) we find that
the last tetrad vector is given by
$$
{e^{4 '}} = {{|\Lambda |}^{-1}} \; \left ( v d t \;
+ \; 8 \beta {\rho ^2} z d \phi \right ) \; \; \; .
$$
Thus the metric is given by
$$
d {s^2} = \left ( 1 \, + \, {\beta ^2} {{\cal E}^2} \right ) \; \left [
{v^2} d {z^2} \; + \; {v^2} d {\rho ^2} \; + \; {\rho ^2} {v^{ - 2}} d
{\phi ^2} \right ]
\; - \; {{\left ( 1 \, + \, {\beta ^2} {{\cal E}^2} \right )}^{-1}}
\; {{\left [ v d t \; + \; 8 \beta z {\rho ^2} d \phi \right ] }^2} \; \; \; .
\eqno({\rm III}.9)
$$

The new Maxwell tensor is given in terms of the new tetrad by
$$
F = 2 {{|\Lambda |}^{- 4}} {v^{-2}} \left [ \left ( 1 \,
+ \, {\beta ^2} {\cal E} \left [ {\cal E} \, + \,
16 {z^2} \right ] \right ) {e^{3 '}} \wedge {e^{2 '}} \;
+ \; 8 {\beta ^2} {\cal E} v z \rho {e^{1 '}} \wedge {e^{ 2 '}} \right ]
$$
$$
+ \; 2 {{|\Lambda |}^{- 4}} {v^{-2}} \beta \left [ \left (
{\cal E} \left [ 1 \, + \, {\beta ^2} {{\cal E}^2} \right ] \,
+ \, 8 {z^2} \left [ {\beta ^2} {{\cal E}^2} \, - \,
1 \right ] \right ) {e^{1 '}} \wedge {e^{4 '}} \;
+ \; 4 v z \rho\left ( 1 \, - \, {\beta ^2}
{{\cal E}^2} \right ) {e^{3 '}} \wedge {e^{4 '}} \right ] \; \; \; .
\eqno({\rm III}.10)
$$
The corresponding electric and magnetic fields are
$$
E =  2 {{|\Lambda |}^{- 4}} {v^{-2}} \beta \left [ \left (
{\cal E} \left [ 1 \, + \, {\beta ^2} {{\cal E}^2} \right ] \,
+ \, 8 {z^2} \left [ {\beta ^2} {{\cal E}^2} \,
- \, 1 \right ] \right ) {e^{1 '}}  \; + \; 4 v z \rho\left ( 1 \,
- \, {\beta ^2} {{\cal E}^2} \right ) {e^{3 '}}  \right ] \; \; \; ,
$$
$$
B = 2 {{|\Lambda |}^{- 4}} {v^{-2}} \left [ \left ( 1 \,
+ \, {\beta ^2} {\cal E} \left [ {\cal E} \, + \, 16
{z^2} \right ] \right ) {e^{1 '}} \; - \; 8 {\beta ^2} {\cal E}
v z \rho {e^{3 '}} \right ] \; \; \; .
\eqno({\rm III}.11)
$$
Note that the electric and magnetic fields each have components in both the
$ z $
and
$ \rho $
directions.  Here again the metric and Maxwell fields reduce to those
of the magnetic universe in the limit as
$ \beta \to 0 $.

\subsection{The $z$ metric}

The steps used to find the new metric using the Killing vector
$ {\bf \partial /\partial z} $
are completely analogous to those used in the case of the
$ {\bf \partial /\partial t} $
Killing vector.  In fact the
$ z $
metric can be found from the
$ t $
metric using the complex coordinate transformation
$ t \to i z , \; z \to i t $.
We will not repeat the derivation; but simply write down the metric:
$$
d {s^2} = \left ( 1 \, + \, {\beta ^2} {{\cal E}^2} \right ) \; \left [ -
{v^2} d {t^2} \; + \; {v^2} d {\rho ^2} \; + \; {\rho ^2} {v^{ - 2}} d
{\phi ^2} \right ]
$$
$$
+ \; {{\left ( 1 \, + \, {\beta ^2} {{\cal E}^2} \right )}^{-1}}
\; {{\left [ v d z \; + \; 8 \beta t {\rho ^2} d \phi \right ] }^2} \; \; \; .
\eqno({\rm III}.12)
$$
where the scalar
$ \cal E $
is here given by
$ {\cal E} = - \left ( {v^2} \, + \, 4 {t^2} \right ) $.

\section {NONSINGULARITY $-$ GEODESICS AND COMPLETENESS}

In this section we demonstrate that these new metrics are nonsingular.
It is clear from the coordinate components that the metrics are smooth for
$ \rho > 0 $.
We now show that the metrics are smooth on the axis as well.  Note that
in general cylindrical coordinates are badly behaved on the axis; so
we must introduce Cartesian coordinates and show
that the corresponding coordinate components of the metric are
smooth.  First consider the case of the
$ \phi $
metric.  Introduce Cartesian coordinates
$ x $
and
$ y $
by
$$
x \equiv \rho \cos \, {\phi \over {1 + {\beta ^2}}} \; \; \; ,
\eqno({\rm IV}.1)
$$
$$
y \equiv \rho \sin \, {\phi \over {1 + {\beta ^2}}} \; \; \; .
$$
Then
$ {\rho ^2} = {x^2} + {y^2} $
so any smooth function of
$ \rho ^2 $
is also a smooth function of
$ x $
and
$ y $.
Using equation (III.3) some straightforward but tedious algebra shows
that the metric in these coordinates is
$$
d {s^2} = \left ( {v^2} + {\beta ^2} \right ) \left [ - \, d {t^2} \,
+ \, d {z^2} \right ] \; + \; {{ \left ( {v^2}
+ {\beta ^2} \right ) }^{- 1}} 8 \beta \left [
2 \beta z {\rho ^2} d {t^2} \, - \, \left ( 1 + {\beta ^2} \right ) \,
d t \, \left ( x d y \, - \, y d x \right ) \right ]
$$
$$
+ \; {{ \left ( {v^2} + {\beta ^2} \right ) }^{- 1}} \left [
{{\left ( 1 + {\beta ^2} \right ) }^2} \left ( d {x^2} \,
+ \, d {y^2} \right ) \; + \; (v + 1 ) \left ( {v^2}
+ 2 {\beta ^2} + 1 \right ) {{\left ( x d x \,
+ \, y d y \right ) }^2} \right ] \; \; \; .
\eqno({\rm IV}.2)
$$
The metric components are all smooth functions of the Cartesian coordinates
($x,y,z,t$).
thus the topology of this spacetime is
$ R^4 $
and the metric is smooth everywhere.

Now consider the
$ t $
metric.  Here we introduce Cartesian coordinates
$ x $
and
$ y $
by
$ x \equiv \rho \cos \phi , \, y \equiv \rho \sin \phi $.
Again
$ {\rho ^2} = {x^2} + {y^2} $
so a smooth function of
$ \rho ^2 $
is a smooth function of
$ x $
and
$ y $.
The metric is given (using equation (III.9) and some straightforward
but tedious algebra) by
$$
d {s^2} = \left ( 1 \, + \, {\beta ^2} {{\cal E}^2} \right )
{v^2} \, d {z^2} \; - \; {v^2} {{\left ( 1 \, + \, {\beta ^2}
{{\cal E}^2} \right ) }^{- 1}} {{\left [ d t \, + \, 8 \beta z
{v^{- 1}} \left ( x d y \, - \, y d x \right ) \right ] }^2}
$$
$$
+ \; \left ( 1 \, + \, {\beta ^2} {{\cal E}^2} \right )
{v^{-2}} \left [ d {x^2} \, + \, d {y^2} \, + \,
(v + 1 ) \left ( {v^2} + 1 \right ) {{\left ( x d x \,
+ \, y d y \right ) }^2} \right ] \; \; \; .
\eqno({\rm IV}.3)
$$
Again the metric components are smooth functions of the coordinates
($x,y,z,t$).
The
$ z $
metric can be obtained from the
$ t $
metric by the transformation
$ t \to i z , \, z \to i t $.
Thus we have also demonstrated that the
$ z $
metric is smooth.

We now study geodesics in the three spacetimes and demonstrate that the
spacetimes are timelike and null geodesically complete
({\it i.e.} nonsingular).  The geodesics will be found using
the Hamilton-Jacobi equations.  We first find a function
$ S $
satisfying
$$
{g^{\mu \nu}} {S_\mu} {S_\nu} \; + \; \kappa = 0
\eqno({\rm IV}.4)
$$
where
$ {S_\mu} \equiv \partial S / \partial {x^\mu} $
and
$ \kappa $
is 0 for null geodesics and 1
for timelike geodesics.  The trajectory of the particle (or light ray)
is then given by solving the equation
$$
{{\dot x}^\mu} = {g^{\mu \nu}} {S_\nu}  \; \; \; .
$$
Here
$ x^\mu $
gives the coordinates of the particle and an overdot denotes derivative with
respect to affine parameter.

\subsection{The $\phi$ metric}

Using the form of the metric we find that the Hamilton Jacobi equation becomes
$$
{S_\rho ^2} \; + {S_z ^2} \; + \; {{{\left ( {v^2} \, + \,
{\beta ^2} \right )}^2} \over {\rho ^2}} \; {S_\phi ^2} \; - \;
{{\left ( 4 \beta z {S_\phi} \,
- \, {S_t}\right ) }^2} \; + \; \kappa \left ( {v^2} \,
+ \, {\beta ^2} \right ) = 0 \; \; \; .
\eqno({\rm IV}.5)
$$
This equation can be separated as follows:
$$
S = \int {\cal Z} ( z ) d z \; + \; \int {\cal R} (\rho ) d \rho \;
+ \; L \phi \; - \; E t \; \; \; .
\eqno({\rm IV}.6)
$$
Here
$ L $
and
$ E $
are constants.  Substituting in the Hamilton-Jacobi equation (IV.5)
we find that the longitudinal and radial momentum functions
$ \cal Z $
and
$ \cal R $
are given by
$$
{\cal Z} = {{\left [ {{\left ( 4 \beta z L \, + \, E \right ) }^2} \;
- \; {U^2} \right ] }^{1/2}} \; \; \; ,
\eqno({\rm IV}.7)
$$
$$
{\cal R} = {{\left [ {U^2} \; - \; {{{\left ( {v^2} \, + \,
{\beta ^2} \right )}^2} \over {\rho ^2}} \; {L^2} \; - \; \kappa
\left ( {v^2} \, + \, {\beta ^2} \right ) \right ] }^{1/2}} \; \; \; .
$$
Here
$ U $
is a separation constant.  The constants
$ - {S_t} = E $
and
$ {S_\phi} =  L $
are conserved quantities related to the Killing vectors
$ \bf \partial /\partial t $
and
$ \bf \partial /\partial \phi $
respectively.  Essentially
$ E $
is the energy of the particle and
$ L $
is its angular momentum.  In the limiting case,
$ \beta = 0 $,
when the solution returns to being that of the seed magnetic universe, the
separation constant
$ U = {{\left [ {E^2} - {{\cal Z}^2} \right ] }^{1/2}} $
is the ``transverse energy''\cite{melvin2}
The presence of a third conserved quantity
$ U $
in this more general case is related to the fact that this
spacetime admits a Killing tensor.  The
Killing tensor
$ K_{\mu \nu} $
is given by
$$
{K_{\mu \nu}} = {{\left ( {v^2} \, + \, {\beta ^2} \right )}^2} \; \left [
{\partial _\mu} z \; {\partial _\nu} z \; - \; {\partial _\mu} t \;
{\partial _\nu} t \right ] \; \; \; .
\eqno({\rm IV}.8)
$$

The geodesic equation then becomes
$$
{\dot \rho } = {{\left [ {U^2} \, {{\left ( {v^2} + {\beta ^2} \right ) }^{-2}}
\; - \; {\rho ^{- 2}} \, {L^2} \; - \; \kappa \, {{\left ( {v^2} +
{\beta ^2} \right ) }^{-1}} \right ] }^{1/2}} \; \; \; ,
\eqno({\rm IV}.9)
$$
$$
{\dot z} = {{\left ( {v^2} + {\beta ^2} \right ) }^{-1}} \,
{{\left [ {{\left ( E \, + \, 4 \beta z L  \right ) }^2} \;
- \; {U^2} \right ] }^{1/2}} \; \; \; ,
$$
$$
{\dot \phi } = {\rho ^{- 2}} \, \left ( {v^2} + {\beta ^2} \right ) \,
L \; - \; 4 \beta z \, {{\left ( {v^2} + {\beta ^2} \right ) }^{-1}} \,
\left ( E \, + \, 4 \beta z L \right ) \; \; \; ,
$$
$$
{\dot t} = {{\left ( {v^2} + {\beta ^2} \right ) }^{-1}} \,
\left ( E \, + \, 4 \beta z L \right ) \; \; \; .
$$
First consider the case
$ L \ne 0 $.
Then
$ \rho $
oscillates between a minimum and a maximum value.  The coordinate
$ z $
is then given by
$$
\int \; {{d \lambda } \over {{v^2} + {\beta ^2}}} \; = \int \;
{{d z} \over {\sqrt {{{\left ( E \, + \, 4 \beta z
L  \right ) }^2} \; - \; {U^2}}}}
\eqno({\rm IV}.10)
$$
where
$ \lambda $
is the affine parameter of the geodesic.  Thus at any finite value of
$ \lambda $
the coordinate
$ z $
remains finite.  It then follows from equations (IV.9) that
$ \phi $
and
$ t $
remain finite for all finite values of
$ \lambda $.
Thus the geodesics are complete.

Now consider the case where
$ L = 0 $.
Then
$ \rho $
is bounded unless
$ \kappa = 0 $.
(When
$ \kappa = L = 0 $
the coordinate
$ \rho $
still remains finite for finite
$ \lambda $).
The coordinate
$ z $
is then given by
$$
z = \int \; d \lambda \; {{\left ( {v^2} + {\beta ^2} \right ) }^{- 1}} \;
{\sqrt {{E^2} - {U^2}}} \; \; \; .
\eqno({\rm IV}.11)
$$
Thus
$ z $
remains finite at finite
$ \lambda $.
Using equations (IV.9) one can then show that
$ \phi $
and
$ t $
also remain finite for finite
$ \lambda $.
Thus the geodesics are complete.  We have thus shown that the
$ \phi $
metric is timelike and null geodesically complete.

\subsection{The $ t $ metric}

Using the form of the metric we find that the Hamilton Jacobi equation becomes
$$
{S_\rho ^2} \; + \; {S_z ^2} \; + \; {{\left ( {v^2} {\rho ^{- 1}} {S_\phi }
\, - \, 8 \beta z \rho v {S_t} \right ) }^2} \; - \;
{{\left ( 1 \, + \, {\beta ^2} {{\cal E}^2} \right ) }^2}
{S_t ^2} \; + \; \kappa  \left ( 1 \, + \, {\beta ^2}
{{\cal E}^2} \right ) {v^2} = 0 \; \; \; .
\eqno({\rm IV}.12)
$$
This equation can be separated as follows:
$$
S = {\cal H} (z,\rho ) \; + \; L \phi \; - \; E t \; \; \; .
\eqno({\rm IV}.13)
$$
Note that
$ \cal H $
cannot be further separated.  This metric has no Killing tensor;
so we have only the two constants of the motion
$ E $
(the energy) and
$ L $
(the angular momentum).

The geodesic equation becomes
$$
{\dot \rho } = {v^{- 2}} {{\left ( 1 \, + \, {\beta ^2}
{{\cal E}^2} \right ) }^{- 1}} {{\cal H}_\rho } \; \; \; ,
$$
$$
{\dot z } = {v^{- 2}} {{\left ( 1 \, + \, {\beta ^2}
{{\cal E}^2} \right ) }^{- 1}} {{\cal H}_z } \; \; \; ,
\eqno({\rm IV}.14)
$$
$$
{\dot \phi} = {{\left ( 1 \, + \, {\beta ^2} {{\cal E}^2} \right ) }^{- 1}} \,
v {\rho ^{- 1}} \, \left ( v {\rho ^{- 1}} L \, + \, 8 \beta z \rho E \right )
$$
$$
{\dot t} = \left ( 1 \, + \, {\beta ^2} {{\cal E}^2} \right )
{v^{- 2}} E  \; - \; {{\left ( 1 \, + \, {\beta ^2}
{{\cal E}^2} \right ) }^{- 1}} \,
8 \beta z \rho \,  \, \left ( v {\rho ^{- 1}} L \,
+ \, 8 \beta z \rho E \right )
$$
Using equations (IV.14) in the Hamilton-Jacobi equation (IV.12) we find
$$
{{\dot \rho }^2} \; + \; {{\dot z}^2} \; - \; {v^{- 4}} \, {E^2} \;
+ \; {v^{- 2}} \, {{\left ( 1 \, + \, {\beta ^2}
{{\cal E}^2} \right ) }^{- 2}} \, {{\left ( v {\rho ^{- 1}} L \,
+ \, 8 \beta z \rho E \right ) }^2} \; + \; \kappa \,
{{\left ( 1 \, + \, {\beta ^2} {{\cal E}^2} \right ) }^{- 1}} \,
{v^{- 2}} = 0 \; \; \; .
$$
It then follows that
$$
{{\dot \rho }^2} \; + \; {{\dot z}^2} \le {v^{- 4}} \, {E^2}
$$
and since
$ v \ge 1 $
we find that
$ \left | {\dot \rho } \right | \le \left | E \right | $
and
$ \left | {\dot z } \right | \le \left | E \right | $.
Therefore
$ \rho $
and
$ z $
remain finite for all finite affine parameter
$ \lambda $.
it then follows from equations (IV.14) that
$ \phi $
and
$ t $
remain finite for all finite
$ \lambda $.
Thus this spacetime is null and timelike geodesically complete.

This method cannot be used to show completeness of the $z$ metric.
However, the fact that the
$z $ and $t$ metrics are related by a
(complex) coordinate transformation leads us to believe that
the $ z $ metric is also null and timelike geodesically complete.

\section{SUMMARY}

The results are summarized in table I.
New Einstein-Maxwell spactimes are generated from the magnetic universe
as a seed metric.  The latter has three Killing vector fields
$ \bf {K_z}, \, {K_\phi}, \, {K_t}$.
The new metrics are found by applying the Ehlers transformation with
the Maxwell-Ernst and Einstein-Ernst potentials to find the new Killing
form and the new tetrad basis forms.
\begin{tabular}{|l|l|l|l|} \hline
New metric & z - metric & $\phi \; $- metric & t - metric \\ \hline
Seed Killing form $K$ & ${K^z} = {v^2} d z $ & ${K^\phi } = \left ({\rho ^2}
/ {v^2} \right )   d \phi $ & ${K^t} = - \; {v^2} d t $ \\
Norm $ f $ & $f = - \, {v^2} $ & $ f = - \, {\rho ^2} /
{v^2} $ & $ f = {v^2} $ \\ \hline
Maxwell potential $\Phi$  & ${\Phi _z} = - \,
2i t$ & ${\Phi _\phi} = - \, 1 / v $ & ${\Phi _t} = 2 i z $ \\ \hline
Einstein potential $\cal E$ & $ {{\cal E}_z} = - \, \left (
{v^2} \, + \, 4 {t^2} \right ) $ & ${{\cal E}_\phi } = - \,
1 / v $ & ${{\cal E}_t} =   {v^2} \, - \, 4 {z^2}  $ \\
1-form $A^E$ & ${A_z ^E} = {{\cal E}_z} d z \, - \, 4 i {\rho ^2}
{v^{- 1}} t d \phi   $ & ${A_\phi ^E} = - \, {v^{- 1}} d \phi \, - \,
2 i z d t $ & $ {A_t ^E} = {{\cal E}_t} d t \, + \, 4 i {\rho ^2}
{v ^{- 1}} z d \phi  $\\
1-form $M^{EE}$ & $ {M^{EE} _z} = {{\cal E}^2 _z} d z $ &
$ {M^{EE} _\phi} = {v^{- 2}} d \phi $ & $ {M^{EE} _t} =
{{\cal E}^2 _t} d t $ \\ \hline
$ {|\Lambda |}^2 $ & $ 1 \, + \, {\beta ^2} {{\cal E}^2 _z} $ &
$ 1 \, + \, {\beta ^2} {v^{ - 2}} $ & $ 1 \, + \, {\beta ^2}
{{\cal E}^2 _t} $ \\ \hline
New tetrad form $e^{ k '} $ & $ {{|\Lambda |}^{- 1}} \left ( v d z \,
+ \, 8 \beta {\rho ^2} t d \phi \right ) $ & $\rho {v^{ -1}}
{{|\Lambda |}^{- 1}} \left ( d \phi \, + \, 4 \beta z
d t \right ) $ & $ {{|\Lambda |}^{- 1}} \left ( v d t \, + \,
8 \beta {\rho ^2} z d \phi \right ) $ \\ \hline
Tetrad forms $ {e^{a '}} \perp {e^{k '}} $ & $ |\Lambda | {e^a} $  &
$ |\Lambda | {e^a} $ & $ |\Lambda | {e^a} $\\ \hline
\end{tabular}

\section{ACKNOWLEDGEMENTS}

We would like to thank Susan Scott for helpful discussions.  This research
was supported in part by NSF grant PHY89-04035 to the Institute for
Theoretical Physics, University of California, Santa Barbara.

\end{document}